# Model for the Fractional Quantum Hall Effect problem

M.I. Dyakonov

*Laboratoire Charles Coulomb, Universite Montpellier 2, CNRS*
*place E. Bataillon, 34095 Montpellier, France*

A simple one-dimensional model is proposed, in which $N$ spinless repulsively interacting fermions occupy $M>N$ degenerate states. It is argued that the energy spectrum and the wavefunctions of this system strongly resemble the spectrum and wavefunctions of 2D electrons in the lowest Landau level (the problem of the Fractional Quantum Hall Effect). In particular, Laughlin-type wavefunctions describe ground states at filling factors $v = N/M = 1(2m + 1)$. Within this model the complimentary wavefunction for $v = 1 - 1/(2m + 1)$ is found explicitly and extremely simple ground state wavefunctions for arbitrary odd-denominator filling factors are proposed.

1. INTRODUCTION

After 20 years of intensive experimental and theoretical studies, our understanding of the Fractional Quantum Hall Effect (FQHE) is far from being satisfactory, although a large progress has been made in this enormously difficult theoretical problem due to the outstanding work in Refs. 1-6. Currently, we are in an awkward position: on the one hand many experimental facts support Jain's idea [3] of composite fermions moving in a reduced effective magnetic field, and this is the only physical description available. On the other hand [7], nobody has really shown theoretically, apart from what may be described as wishful thinking, the existence of composite fermions, as (quasi) free particles. Moreover, this concept does not provide answers to a number of simple and fundamental questions; it does not even explain what a composite fermion is.

The basic problem of FQHE can be formulated as follows. We have $N$ two-dimensional electrons in the lowest Landau level containing $M > N$ degenerate states. The electrons may be regarded as fully spin-polarized (or spinless) fermions confined to the lowest Landau level and higher Landau levels are irrelevant. We are interested in finding the energy spectrum and the wavefunctions of this system for arbitrary filling $v = N/M < 1$.

After introducing natural units of length and energy we are left with a dimensionless problem of diagonalizing a huge numerical matrix with a single dimensionless parameter, $v$. If $v$ were very small, one could use it as a small parameter, and indeed there is theoretical, as well as experimental, evidence that for small enough $v$ a Wigner crystal is formed. However the most interesting phenomena occur when $v$ is *not* very small. In this case our purely numerical problem appears to be intractable theoretically, since no approximation can be justified, and we have to rely on numerical calculations. It would be only of minor interest to know the exact numerical values of the ground state and excited state energies, if we were not aware of the totally unexpected *experimental* fact that something special happens at rational values of $v$, namely that gaps in the excitation spectrum appear when $v = p/q$ with $q$ odd. This result is not understood theoretically, except for $v=1/(2m+1)$.

Laughlin [1] gave a simple and clear answer to the question: *why does something special happen at $v = 1/(2m+1)$?* −Because, for this filling a ground state wavefunction may be constructed, which goes to zero at $z_i \rightarrow z_j$ like $(z_i - z_j)^{2m+1}$, faster than for neighbouring values of $v$, thus minimizing the electron repulsion energy ($z_i$ is the complex coordinate of the $i$-th electron). This circumstance is responsible for the gap in the energy spectrum.

While this idea certainly gave a clue to understanding the FQHE, some important questions remain unanswered. One of them concerns the $v=2/3$ state (and, generally, all the

$v=1-1/(2m+1)$ states). Because of the electron-hole symmetry, this state, $\Psi_{2/3}$, can be regarded as the $v=1/3$ *hole* state described by the Laughlin wavefunction, $\Psi_{1/3}$, depending on the coordinates of *N holes* in a completely filled Landau level. The physical properties should be (and, in fact, are) quite similar to those at $v=1/3$. Suppose, however, that one wants to have a look at the $\Psi_{2/3}$ function written in terms of *2N electron* coordinates. To do this, one must (i) write down the Laughlin function $\Psi_{1/3}$ as a superposition of $N \times N$ determinants involving one-particle hole wavefunctions, and (ii) leaving the coefficients in the superposition unchanged, replace each determinant by its complimentary $2N \times 2N$ electron determinant. The resulting unwieldy expression, which nobody knows how to write down explicitly, will represent the $v=2/3$ ground state, $\Psi_{2/3}$. It will go to zero at $z_i \to z_j$ as $(z_i - z_i)$, just like any antisymmetric function, and we will hardly be able to understand why this function should minimize the interaction energy! This shows the existence of wavefunctions that are as good as the Laughlin function, but which do not have higher order zeros when the electron coordinates coincide. In this sense the $\Psi_{2/3}$ function resembles the wavefunctions for other rational fillings, such as $\Psi_{2/5}$. It remains an open question, what are the relevant properties of these ground state wavefunctions, and this is a clear signal that our understanding is not complete.

There is also a simple, but theoretically important, question: are peculiarities at rational filling factors specific for the FQHE, or will they exist in other situations, when one has *M* degenerate quantum states partially filled by *N* interacting fermions (with repulsive interaction)? Many such models can be proposed, but probably the simplest one is the following one-dimensional problem.

## 2. MODEL

Consider *M* degenerate one-particle states on a circle:

$$\psi_k(\varphi) = \frac{1}{\sqrt{2\pi}} \exp(ik\varphi), \qquad k = 0..M-1. \qquad (1)$$

There are $N<M$ spinless fermions, which repel each other via some pair potential $U(\varphi_i - \varphi_j)$. Find energy spectrum for a given *v*. Certainly the problem is somewhat artificial, and of course no Hall effect will exist. However, the question we are interested in is whether the energy spectrum for this system resembles that for the true FQHE system. Will there be gaps in the excitation spectrum at $v=p/q$ with *q* odd, gapless states at $v=1/2m$, and fractionally charged quasi-particles?

It should be stressed that this model, designed as a caricature of the FQHE system, is very different from a model, in which fermions can occupy any of *M fixed* sites on a circle. Although Wannier-type localized functions can be introduced by the relation

$$\Phi_s(\varphi) = \Phi(\varphi - \frac{2\pi}{M}s) = \frac{1}{\sqrt{M}} \sum_{k=0}^{M-1} \psi_k(\varphi) \exp(-\frac{2\pi i}{M}ks), \qquad s = 0..M-1, \qquad (2)$$

the choice of the localization sites is arbitrary (a different, but equivalent basic set can be obtained by shifting all the sites $\varphi_s = 2\pi s/M$ by an arbitrary angle).

A crystal-like many-particle state may be constructed as a suitable determinant of these functions, which presumably will give the ground state for small enough *v*. However, if

$v=1/(2m+1)$ is not very small, a Laughlin-type function with uniform density should be preferable. This function may be readily written as

$$\Psi_{1/(2m+1)}(\varphi_1...\varphi_N) = A \prod_{1\leq i<j\leq N}\left[\exp(i\varphi_i)-\exp(i\varphi_j)\right]^{2m+1}, \quad A^2 = \frac{[(2m+1)!]^N}{(2\pi)^N[(2m+1)N]!}. \quad (3)$$

Note that, in contrast to the case of the original Laughlin function, for our problem the normalization constant $A$ is found analytically, the corresponding normalization integral having been calculated by Dyson [8, 9]. Exactly following Laughlin, quasi-holes and quasielectrons may be introduced, and the same arguments will lead us to the conclusion that gaps in the excitation spectrum should appear for $v=1/(2m+1)$. Thus it seems that, at least for the Laughlin states, there is no great difference between our model and the true FQHE system.

Interestingly, within our model a simple answer can be given to the question concerning the complimentary wavefunction at $v=1-1/(2m+1)$, which was discussed above. It may be proved [10] that, given the Laughlin-type function for $(2m+1)N=M$, Eq. (3), the complimentary wavefunction for $M-N$ particles at $v=2m/(2m+1)$, derived as indicated above, has a similar form:

$$\Psi_{1-1/(2m+1)}(\varphi_1...\varphi_{M-N}) = B \prod_{1\leq i<j\leq M-N}\left[\exp(i\varphi_i)-\exp(i\varphi_j)\right]^{2m+1}, \quad (4)$$

with a known normalization constant $B$. The rhs of Eq. (4) contains powers of each $\exp(i\varphi_i)$ up to $(2m+1)(M-N-1) = 2m(M-1)$ which is greater than $M$. *These powers should be taken modulo $M$,* and this is a non-trivial *exact* result reminiscent of Jain's projection procedure [3]. Because of the modulo $M$ rule the function in Eq. (4) has only simple zeros when $\varphi_i \to \varphi_j$.

Using an expansion with respect to the basic set in Eq. (2), this result may be rewritten in a form, in which the modulo $M$ rule is applied automatically. In this basis, Eq. (4) becomes:

$$\Psi_{1/(2m+1)}(\varphi_1...\varphi_N) = \sum_{(s)} C(s_1...s_N) \Phi_{s_1}(\varphi_1)...\Phi_{s_N}(\varphi_N), \quad (5)$$

where the sum is over all $s_i$ from 0 to $M-1$ and the coefficients $C$ (which in fact give the same wavefunction in a new representation) are given by:

$$C(s_1...s_N) = A \prod_{1\leq i<j\leq N}(\omega^{s_i}-\omega^{s_j})^{2m+1}, \quad A^2 = \frac{[(2m+1)!]^N}{(M)^N[(2m+1)N]!}, \quad (6)$$

where $\omega$ is root of unity: $\omega = \exp(2\pi i/M)$. Then it can be proven that the complimentary wavefunction,

$$\Psi_{1-1/(2m+1)}(\varphi_1...\varphi_{M-N}) = \sum_{(s)} D(s_1...s_{M-N}) \Phi_{s_1}(\varphi_1)...\Phi_{s_{M-N}}(\varphi_{M-N}), \quad (7)$$

has coefficients $D$, which have an appearance quite similar to Eq. (6):

$$D(s_1...s_{M-N}) = B \prod_{1\leq i<j\leq M-N}(\omega^{s_i}-\omega^{s_j})^{2m+1}. \quad (8)$$

In this expression powers of $\omega^{s_i}$ higher than $M-1$ are automatically reduced to the interval $[0, M-1]$. The explicit form of the complimentary wavefunction given by Eq. (4), or equivalently, by Eqs. (7), (8), is a consequence of the following

<u>Theorem</u> [10]. Let $s_i$ be an arbitrary set of $N$ integers from 0 to $M-1$, and let $p_i$ be a complimentary set of $M-N$ integers (for example, if $M=5$, $N=2$, and $s_1=1$, $s_2=3$, then $p_1=0$, $p_2=2$, and $p_3=4$), then

$$\frac{1}{\sqrt{M^N}} \prod_{1 \leq i < j \leq N} \left|\omega^{s_i} - \omega^{s_j}\right| = \frac{1}{\sqrt{M^{M-N}}} \prod_{1 \leq i < j \leq M-N} \left|\omega^{p_i} - \omega^{p_j}\right|, \quad \omega = \exp\left(\frac{2\pi i}{M}\right). \quad (9)$$

It is aesthetically pleasing that Eqs. (6) and (8) have exactly the same form, and this must have some profound reason which is not yet understood. One is tempted to extrapolate this result for other odd-denominator fillings and suggest expressions like Eqs. (5, 6) as the ground state wavefunctions for *all* $v=p/q$ with $q=2m+1$. Because of Eq. (9), this conjecture is self-consistent, in the sense that *if* it is true for some $v$, it is also true for filling $1-v$.

It will be of considerable interest if somebody would study this simple model numerically. If, as I believe, the results for odd and even denominator fractions will be quite similar to those for interacting electrons in the lowest Landau level, it could shed light on the true FQHE problem.

What about composite fermions? In my one-dimensional model there is no magnetic field, no fluxes to "attach" to electrons, and no loops to carry electrons around. Clearly, for this model the language conventionally adopted by the FQHE theorists, as well as the way of thinking, should be strongly modified, and probably the same is true for the real FQHE problem.

The properties of the energy spectrum at odd-denominator filling factors, which are responsible for the FQHE, are not specific for 2D electrons in a strong magnetic field, but should exist whenever $N$ fermions occupy $M>N$ initially degenerate states, if the interaction is repulsive.